\documentclass[pdflatex,sn-mathphys-num]{sn-jnl}% Math and Physical Sciences Numbered Reference Style
\usepackage{graphicx}%
\usepackage{multirow}%
\usepackage{amsmath,amssymb,amsfonts}%
\usepackage{amsthm}%
\usepackage{mathrsfs}%
\usepackage[title]{appendix}%
\usepackage{xcolor}%
\usepackage{textcomp}%
\usepackage{manyfoot}%
\usepackage{booktabs}%
\usepackage{algorithm}%
\usepackage{algorithmicx}%
\usepackage{algpseudocode}%
\usepackage{listings}%
\theoremstyle{thmstyleone}%
\theoremstyle{thmstyletwo}%

\theoremstyle{thmstylethree}%

\begin{document}

\title[]{Development and demonstration of the Korea ALICE Telescope using electron beams at KEK PF-AR}

\author[1]{\fnm{Jiyoung} \sur{Kim}}\email{jiyoung.kim@cern.ch}
\author[1,2]{\fnm{Meike} \sur{Danisch}}\email{meike.charlotte.danisch@cern.ch}
\author[2]{\fnm{Sungwoon} \sur{Choi}}
\author[5]{\fnm{Tatsuya} \sur{Chujo}}
\author[6]{\fnm{Taku} \sur{Gunji}}
\author[3]{\fnm{Yoonha} \sur{Hong}}
\author[2]{\fnm{Hangil} \sur{Jang}}
\author[4]{\fnm{Towa} \sur{Katsuno}}
\author[6]{\fnm{Ryotaro} \sur{Kohara}}
\author[1]{\fnm{MinJung} \sur{Kweon}}
\author*[3]{\fnm{Sanghoon} \sur{Lim}}\email{shlim@yonsei.ac.kr}
\author[5]{\fnm{Inaba} \sur{Motoi}}
\author[6]{\fnm{Hikari} \sur{Murakami}}
\author[5]{\fnm{Hanseo} \sur{Park}}
\author[5,7]{\fnm{Jonghan} \sur{Park}}
\author[5]{\fnm{Shingo} \sur{Sakai}}
\author[5]{\fnm{Daito} \sur{Shibata}}
\author[4]{\fnm{Reita} \sur{Wada}}
\author[2]{\fnm{Kyungrim} \sur{Woo}}
\author[4]{\fnm{Yorito} \sur{Yamaguchi}}
\author[1]{\fnm{Seunghwan} \sur{Yang}}
\author[2]{\fnm{In-Kwon} \sur{Yoo}}
\author[8]{\fnm{Miljenko} \sur{Suljic}}
\author[9]{\fnm{Serhiy} \sur{Senyukov}}
\author[10]{\fnm{Giacomo} \sur{Contin}}
\author[11,12]{\fnm{Domenico} \sur{Colella}}
\author[6]{\fnm{Bong-Hwi} \sur{Lim}}
\author[13]{\fnm{Mihail-Bogdan} \sur{Blidaru}}
\author[14]{\fnm{Berkin} \sur{Ulukutlu}}

\affil[1]{\orgdiv{Department of Physics}, \orgname{Inha University}, \orgaddress{\city{Incheon}, \postcode{22212}, \country{South Korea}}}
\affil[2]{\orgdiv{Department of Physics}, \orgname{Pusan National University}, \orgaddress{\city{Busan}, \postcode{46241}, \country{South Korea}}}
\affil[3]{\orgdiv{Department of Physics}, \orgname{Yonsei University}, \orgaddress{\city{Seoul}, \postcode{03722}, \country{South Korea}}}
\affil[4]{\orgdiv{Physics Program and International Institute for Sustainability with Knotted
Chiral Meta Matter}, \orgname{Hiroshima University}, \orgaddress{\city{Hiroshima}, \postcode{739-8526}, \country{Japan}}}
\affil[5]{\orgdiv{Institute of Pure and Applied Sciences}, \orgname{University of Tsukuba}, \orgaddress{\city{Tsukuba}, \postcode{305-8571}, \country{Japan}}}
\affil[6]{\orgdiv{Center for Nuclear Study}, \orgname{University of Tokyo}, \orgaddress{\city{Tokyo}, \postcode{113-0033}, \country{Japan}}}
\affil[7]{\orgdiv{Division of Science Education}, \orgname{Jeonbuk National University}, \orgaddress{\city{Jeonju}, \postcode{54896}, \country{South Korea}}} 

\affil[8]{\orgname{European Organisation for Nuclear Research (CERN)}, \orgaddress{\city{Geneva}, \postcode{1211}, \country{Switzerland}}}
\affil[9]{\orgdiv{CNRS, IPHC UMR 7178}, \orgname{Universit\'e de Strasbourg}, \orgaddress{\city{Strasbourg}, \postcode{67037}, \country{France}}}
\affil[10]{\orgdiv{Dipartimento di Fisica dell'Universit\`a and Sezione INFN}, \orgname{University of Trieste}, \orgaddress{\city{Trieste}, \postcode{34127}, \country{Italy}}}
\affil[11]{\orgdiv{Department of Physics}, \orgname{University of Bari}, \orgaddress{\city{Bari}, \postcode{70126}, \country{Italy}}}
\affil[12]{\orgname{INFN, Sezione di Bari}, \orgaddress{\city{Bari}, \postcode{70126}, \country{Italy}}}
\affil[13]{\orgname{GSI Helmholtzzentrum f\"ur Schwerionenforschung GmbH}, \orgaddress{\city{Darmstadt}, \postcode{64291}, \country{Germany}}}
\affil[14]{\orgname{Technische Universität München}, \orgaddress{\city{Munich}, \postcode{80333}, \country{Germany}}}

%\author*[1,2]{\fnm{First} \sur{Author}}\email{iauthor@gmail.com}



%\affil*[1]{\orgdiv{Department}, \orgname{Organization}, \orgaddress{\street{Street}, \city{City}, \postcode{100190}, %\state{State}, \country{Country}}}

%\affil[2]{\orgdiv{Department}, \orgname{Organization}, \orgaddress{\street{Street}, \city{City}, \postcode{10587}, \state{State}, \country{Country}}}

%\affil[3]{\orgdiv{Department}, \orgname{Organization}, \orgaddress{\street{Street}, \city{City}, \postcode{610101}, \state{State}, \country{Country}}}


\abstract{
The development of ultra-low-mass, high-precision vertex detectors is a key requirement for future collider experiments and motivates extensive research and development of novel silicon tracking technologies. In this work, we present the development and beam-test demonstration of the Korea ALICE Telescope (KATS), a silicon-tracking telescope designed to support R\&D on next-generation cylindrical vertex detectors, such as the proposed ALICE ITS3 upgrade. The telescope consists of six ALPIDE Monolithic Active Pixel Sensors (MAPS) used as reference tracking planes, a bent ALPIDE sensor serving as the device under test, and a scintillating-fiber-based trigger system, all housed in a light-tight modular enclosure. This setup enables precise track reconstruction and detailed performance studies of both planar and curved silicon sensors.
Beam tests were carried out using high-energy electron beams at the KEK Photon Factory Advanced Ring (PF-AR). The telescope system operated stably under realistic beam conditions, and its tracking performance was successfully validated. The bent ALPIDE sensor was operated at a bending radius of approximately 18 mm, consistent with ITS3's design goals, without any observable degradation in detection performance. The measured results confirm that the KATS provides a versatile and reliable platform for studies of curved MAPS technologies, alignment precision, and tracking performance.
These results provide important experimental validation of key technologies for future low-mass cylindrical silicon vertex detectors and establish KATS as a valuable facility for ongoing and future detector R\&D.
}




\maketitle

\section{Introduction}\label{sec1}

Silicon pixel detectors have become an indispensable component of modern high-energy and nuclear physics experiments, providing the precise spatial information required for charged-particle tracking and vertex reconstruction near the interaction point. Their ability to resolve displaced decay vertices is particularly critical for studying heavy-flavor hadrons and other short-lived particles, which serve as key probes of both Standard Model processes and potential new-physics signatures. Driven by these demands, silicon vertex and tracking systems are now central to the LHC experiments and to the design of future facilities such as the Electron-Ion Collider and proposed next-generation colliders.

The performance of a vertex detector is governed primarily by three factors: the radial distance of the innermost layer from the interaction point, the material budget of the detector layers, and the intrinsic spatial resolution of the sensor technology. A smaller radius and lower material budget directly improve impact parameter resolution, particularly for low-momentum tracks. Achieving both simultaneously, however, is challenging because conventional barrel geometries rely on staves arranged in concentric cylindrical layers, whose mechanical support structures and cooling services impose practical lower limits on the material budget, while the stave geometry itself constrains the achievable innermost radius.

The ALICE Inner Tracking System 2 (ITS2) represents the current state of the art in low-mass vertex detector design, achieving an innermost layer radius of approximately 23 mm with a material budget of about 0.3\% of a radiation length ($X_{0}$) per layer~\cite{ALICE:2013_ITS2TDR}. Despite these design achievements, more than 80\% of the total material budget comes from mechanical support structures and services rather than the silicon sensors themselves, illustrating the intrinsic limitations of stave-based detector geometries.

A promising strategy to overcome these limitations is to realize truly cylindrical detection layers composed of wafer-scale, ultra-thin silicon sensors. Such a design allows detector layers to be placed closer to the beam pipe while largely eliminating the need for conventional support structures, thereby reducing the material budget to nearly that of the sensor itself. This concept forms the basis of the proposed ALICE ITS3 upgrade~\cite{ALICE:2019_ITS3LOI, ALICE:2024_ITS3TDR}, which envisions three layers of curved, wafer-scale Monolithic Active Pixel Sensors arranged in perfectly cylindrical geometry, with the innermost layer positioned at a radial distance of only 18 mm from the interaction point. The ITS3 design is expected to deliver substantial improvements in vertex resolution, tracking efficiency, and low-$p_{\mathrm{T}}$ performance.

A key milestone toward realizing such cylindrical detectors is experimentally demonstrating that Monolithic Active Pixel Sensors can be bent to small radii without degrading their electrical functionality or tracking performance. Initial proof-of-concept studies by the ALICE Collaboration using bent ALPIDE sensors—the MAPS technology employed in ITS2—have demonstrated mechanical feasibility and validated electrical and tracking performance in test-beam measurements~\cite{ALICE:2022_firstBent}. Independent reproduction and systematic follow-up studies at dedicated test-beam facilities are, however, essential to consolidate these findings and support the broader R\&D program toward ITS3 and future curved-MAPS detectors.

Most curved-MAPS test-beam activities to date have been carried out at European facilities such as DESY~\cite{ALICE:2025_secondBent} and the CERN PS/SPS, whose beam time is in high demand across many detector R\&D programs. Establishing complementary test-beam capabilities in Asia would therefore broaden access to beam time and foster active collaboration across regions. Motivated by this, the Korea ALICE group has developed the Korea ALICE Telescope (KATS), a silicon tracking telescope designed to support R\&D for next-generation low-mass cylindrical vertex detectors, with an initial focus on ITS3-related technologies. KATS integrates six ALPIDE sensors as reference tracking planes, a bent ALPIDE sensor as the device under test, and a scintillating-fiber-based trigger system within a modular, light-tight enclosure, and was commissioned using high-energy electron beams at the KEK Photon Factory Advanced Ring (PF-AR)~\cite{KEK_PFAR:2023smx}. In this paper, we describe the design, construction, and beam-test commissioning of KATS, and present an independent validation of bent ALPIDE performance, thereby establishing KATS as a flexible platform for ongoing and future silicon tracking detector R\&D.

\section{Telescope configuration}
The Korea ALICE Telescope (KATS) comprises four principal subsystems: six ALPIDE sensors serving as reference tracking planes, a device under test (DUT)---implemented as a bent ALPIDE sensor in the present study---a scintillating-fiber-based trigger system, and a dedicated telescope enclosure. All detector elements are mounted on an optical breadboard inside the enclosure, ensuring precise alignment and stable positioning throughout beam-test operation. The following subsections describe each subsystem in detail.

\subsection{ALPIDE reference planes}

\begin{table}[htb]
    \centering
    \caption{Specifications of the ALPIDE sensor~\cite{MAGER2016434}.}
    \begin{tabular}{ll}
    \hline
    Parameter & Value \\
    \hline
    Chip dimensions & 15 mm × 30 mm \\
    Pixel pitch & 26.88 $\mu$m × 29.24 $\mu$m \\
    Pixel matrix & 512 × 1024 \\
    Sensor thickness & 50 $\mu$m \\
    Detection efficiency & $>$ 99$\%$ \\
    Spatial Resolution & 5 $\mu$m \\
    Fake-hit rate & 10$^{-6}$ pixel$^{-1}$event$^{-1}$ \\
    Typical frame readout time & 10 $\mu$s \\
    \hline
    \end{tabular}
    \label{tab:alpide_specs}
\end{table}

ALPIDE is a Monolithic Active Pixel Sensor (MAPS) fabricated in a 180 nm CMOS imaging process, originally developed for the ALICE ITS2 upgrade; its key specifications are summarized in Table~\ref{tab:alpide_specs}. ALPIDE was selected for the KATS reference planes because of its well-established performance characteristics and its extensive characterization within the ALICE collaboration. Employing the same sensor type for both the reference planes and the DUT further simplifies the data acquisition and reconstruction chain, allowing all detectors to be read out within a common EUDAQ-based framework.

Three reference planes are placed upstream of the DUT and three downstream, with the DUT positioned at the geometric center of the telescope for optimal track extrapolation. This symmetric arrangement enables high-precision reconstruction of incident particle tracks and accurate interpolation of the track position at the DUT location, providing the essential basis for DUT detection efficiency and spatial resolution measurements. All reference-plane sensors operate at a reverse-substrate bias voltage of $-3$ V, chosen to maximize the detection efficiency of each plane and minimize plane-to-plane performance variations, ensuring uniform and stable track reconstruction across the telescope.

\begin{figure}
    \centering
    \includegraphics[width=0.5\linewidth]{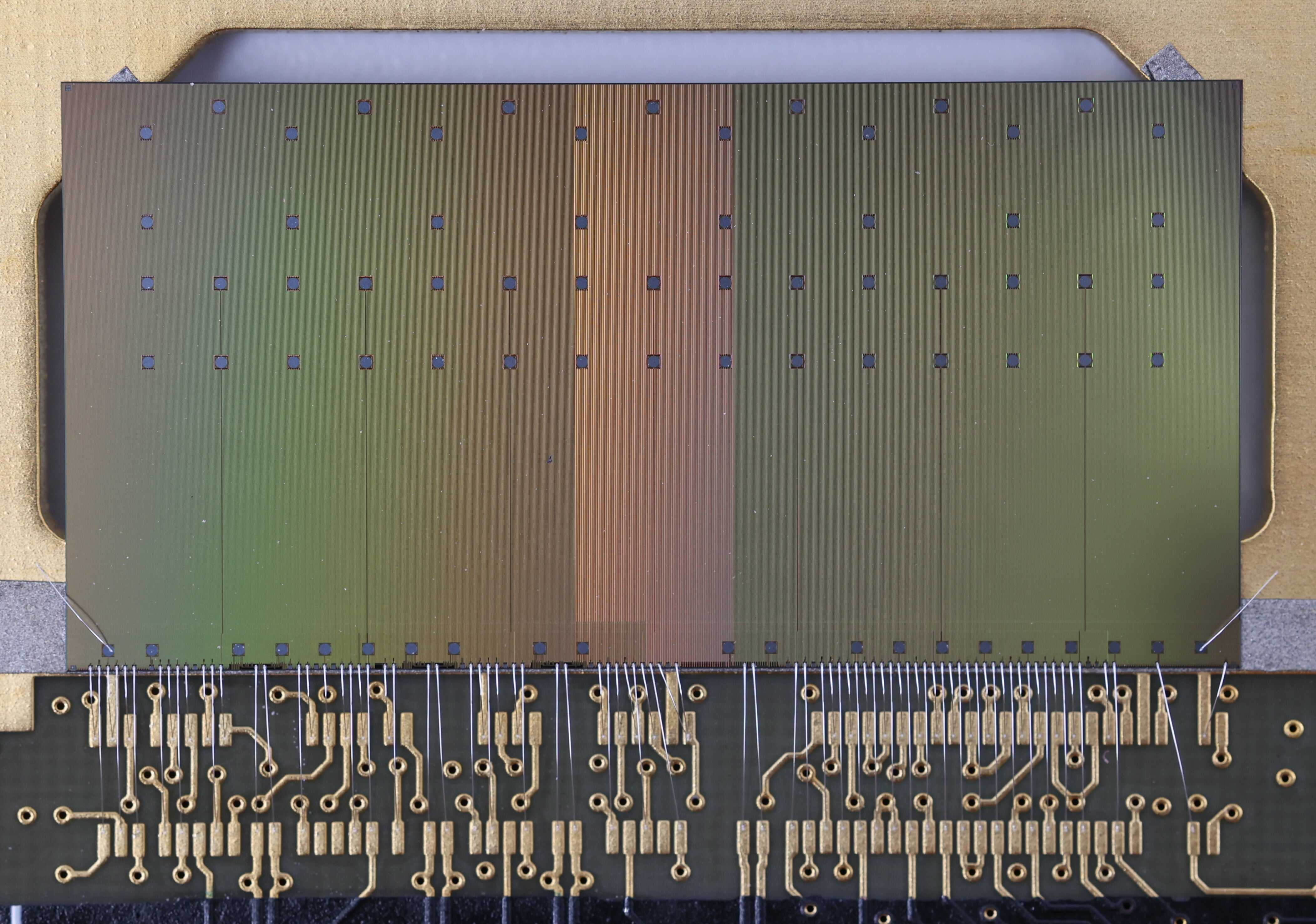}
    \caption{Photograph of the ALPIDE chip assembled on the carrier board.}
    \label{fig:alpide-chip-pic}
\end{figure}

\subsection{Bent ALPIDE}

\begin{figure}
    \centering
    \includegraphics[width=0.85\linewidth]{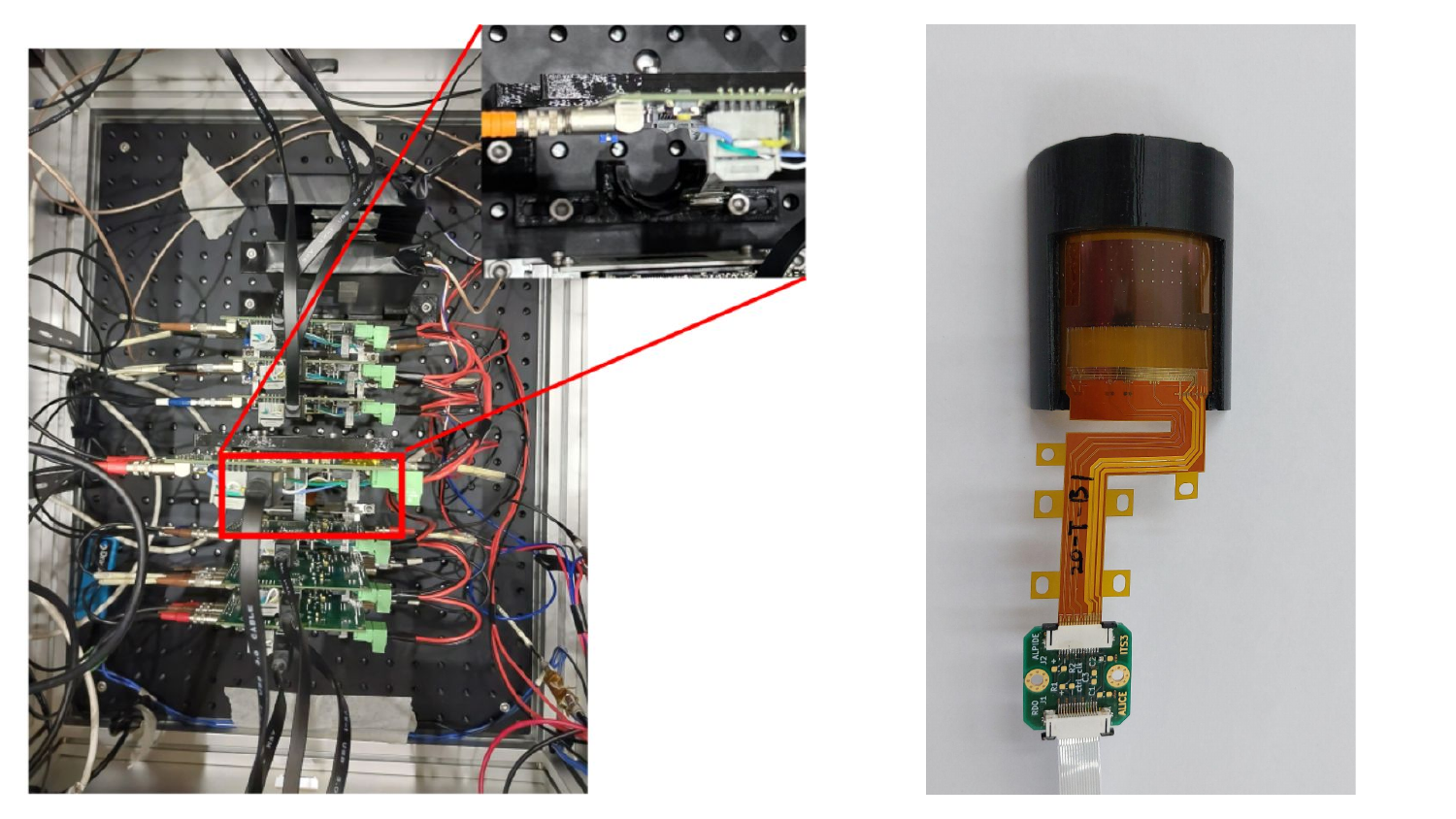}
    \caption{Photograph of the detector planes inside the telescope box, including the bent ALPIDE layer.}
    \label{fig:telescope-bentALPIDE-pic}
\end{figure}

The bent ALPIDE sensor serves as the device under test (DUT) in the present beam test. ALPIDE was chosen for this bending study because its performance in the planar configuration is thoroughly characterized from ITS2 operation, enabling a direct, controlled comparison between planar and curved sensor geometries. In this study, we aim to reproduce and independently validate the results of the earlier ALICE proof-of-concept studies~\cite{ALICE:2022_firstBent, ALICE:2025_secondBent} using the newly developed KATS, thereby demonstrating both the performance of curved ALPIDE sensors at ITS3-relevant bending radii and the capability of KATS as a platform for curved-MAPS R\&D.

The bending procedure developed for this work differs from the original CERN approach, in which the sensor is bent before wire bonding. The flexible printed circuit board (FPC), produced by MEMSPACK\footnote{MEMSPACK Co., Ltd., Republic of Korea} based on design files provided by CERN, incorporates a dedicated rectangular Kapton mounting region for the ALPIDE chip. The chip is first attached to this region with epoxy adhesive and subsequently wire-bonded to the FPC. Performing wire bonding on the still-planar assembly improves handling stability and reduces the risk of mechanical failure during the subsequent bending step. The 3D-printed bending guide was designed with an opening behind the active sensor area to minimize the material budget traversed by particles during beam tests while still providing sufficient mechanical support around the sensor edges. The bending itself is performed with a 3D-printed guide that uniformly conforms the combined chip–FPC assembly to a predefined cylindrical frame via a sliding mechanism.

The resulting bending radius was measured with a VR-6200 three-dimensional optical profilometer and found to be 17.93 mm, in excellent agreement with the ITS3 target radius of 18 mm adopted at the time of this study. The most recent ITS3 R\&D iteration has updated this target to 19 mm, which remains compatible with the demonstrated bending capability.

\subsection{Trigger detector}\label{subsec:trigger detector} 

\begin{figure}[htb]
    \centering
    \includegraphics[width=0.37\textwidth]{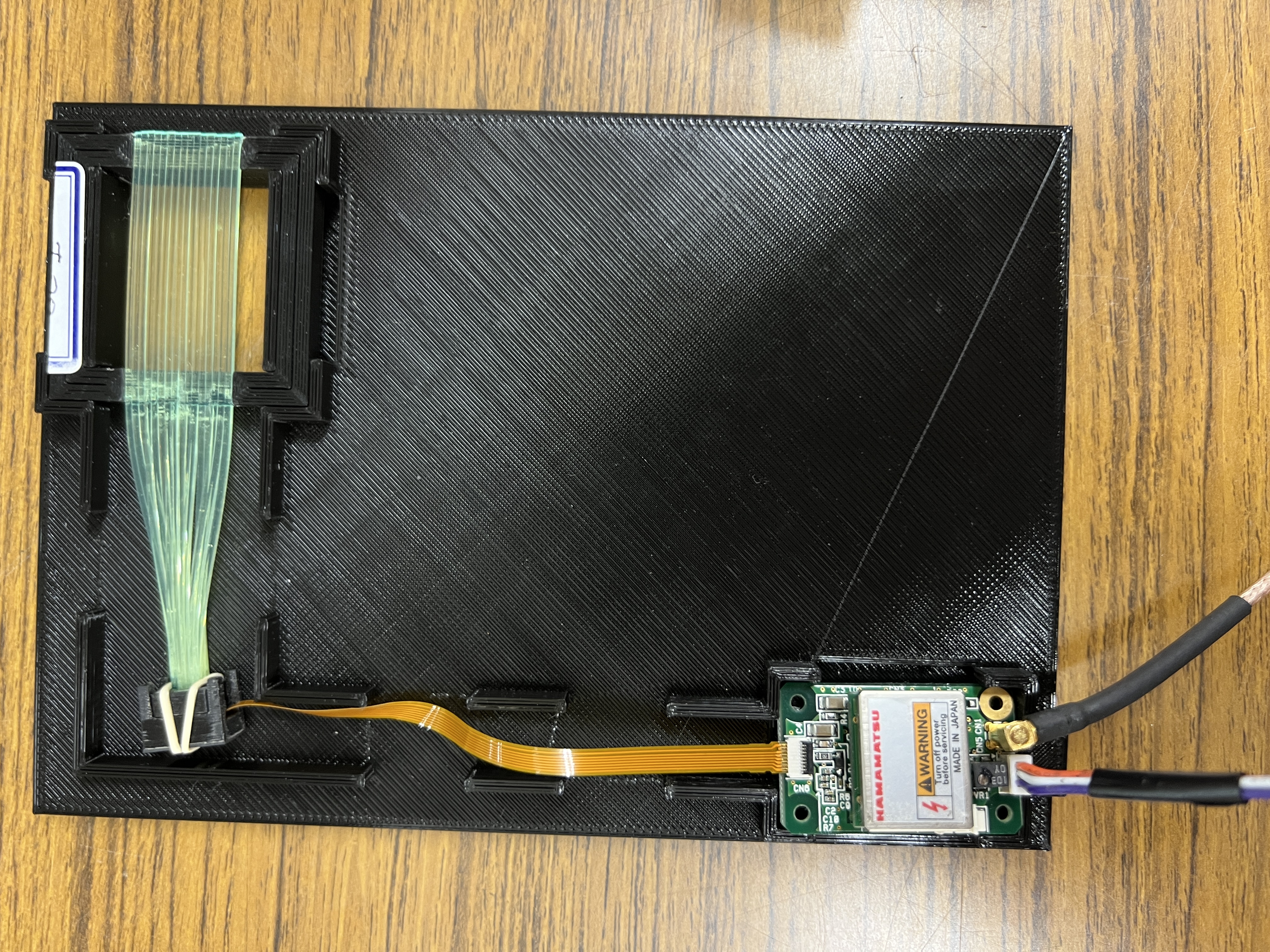}
    \includegraphics[width=0.55\textwidth]{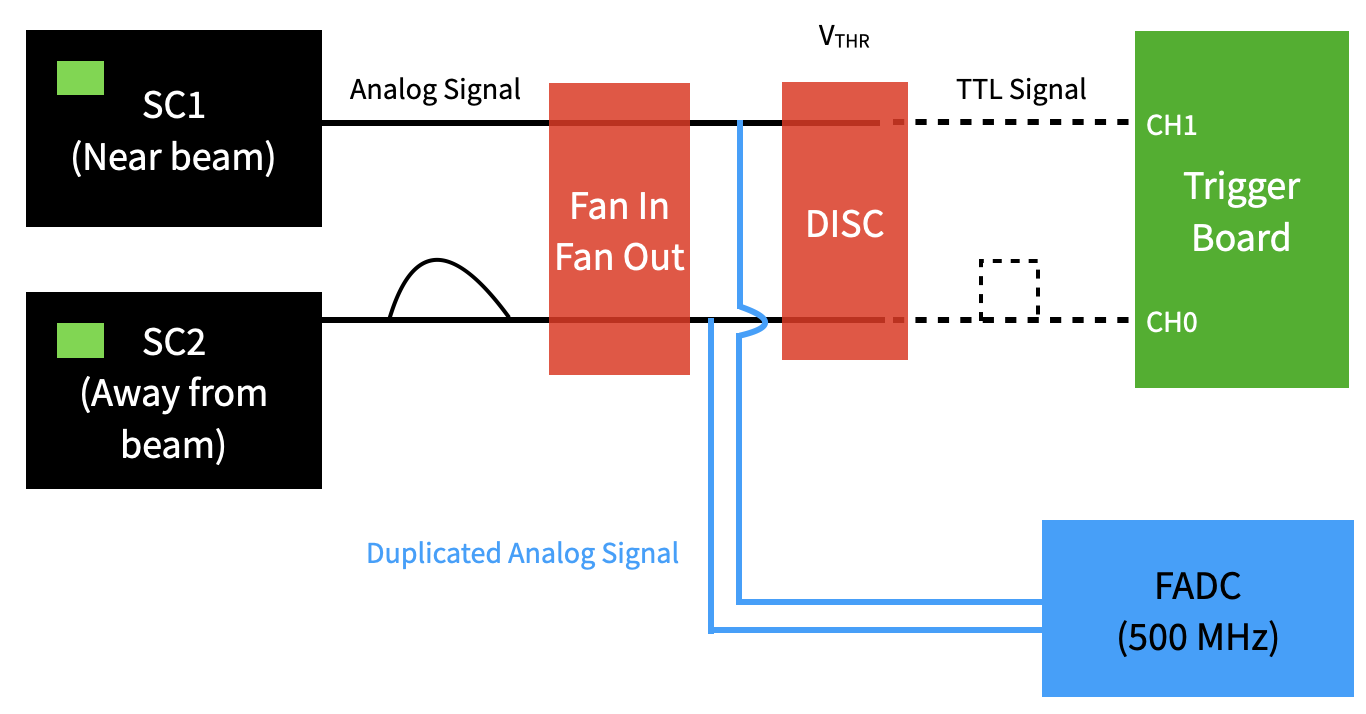}
    \caption{Photographs of the scintillating fiber trigger modules (left) and DAQ/trigger logic schematic (right).}
    \label{fig:trigger}
\end{figure} 

The KATS trigger system is based on two scintillating-fiber\footnote{BCF-20 green fiber from Saint-Gobain} trigger modules, installed side by side downstream of the ALPIDE sensor planes along the beam direction. Each module consists of 16 fibers with a cross-sectional area of 1 mm$^{2}$ per fiber, arranged to fully cover the active area of a single ALPIDE sensor and thereby ensure uniform trigger acceptance across the telescope (left panel of Fig.~\ref{fig:trigger}). The 16 fibers of each module are bundled in a $4\times4$ configuration and optically coupled at one end to the active surface of a single Multi-Pixel Photon Counter (MPPC)\footnote{Hamamatsu C13367-3050EA}, which converts the scintillation light into an electrical signal. To minimize light-induced noise on the MPPCs, each module is assembled inside a dedicated 3D-printed black housing, and the entire module is additionally wrapped in a black light-shielding sheet.

The analog signals from the MPPCs are processed by discriminators, which produce digital logic pulses when the pulse amplitude exceeds a user-configurable threshold. The discriminator outputs are combined in a trigger board that generates the final trigger signal for the telescope (right panel of Fig.~\ref{fig:trigger}). The initial design foresaw a coincidence between two modules placed upstream and downstream of the sensor planes to suppress accidental triggers from MPPC dark counts and other beam-uncorrelated sources. However, the final operational configuration used OR logic between the two modules, both installed downstream, to mitigate the impact of local trigger inefficiencies; the rationale for this change is discussed in \textbf{Section 3.1}.

In parallel, the analog signals from the MPPCs are digitized and recorded using a Flash ADC DAQ system\footnote{NKFADC500 from NOTICE}, preserving full-waveform information for offline analyses of trigger-pulse amplitudes and timing. These waveforms provide direct access to the beam structure and enable detailed performance studies of the trigger system during data-taking.

\subsection{Telescope box} 

\begin{figure}[htb]
    \centering
    \includegraphics[width=0.467\textwidth]{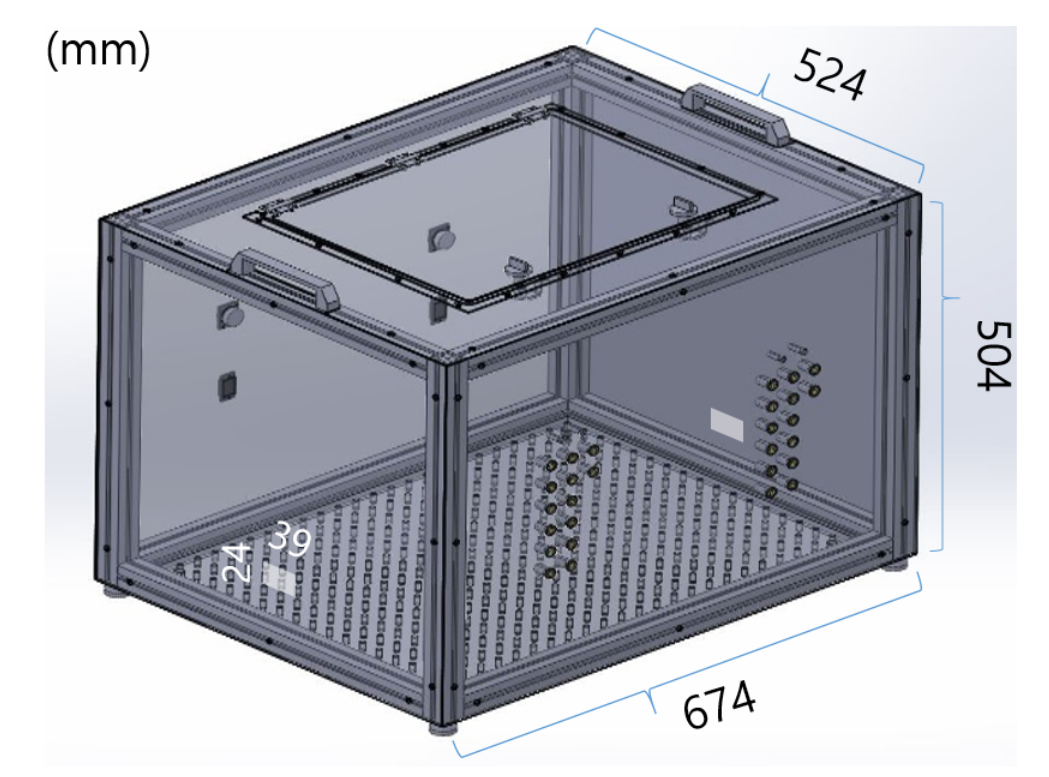}
    \includegraphics[width=0.4\textwidth]{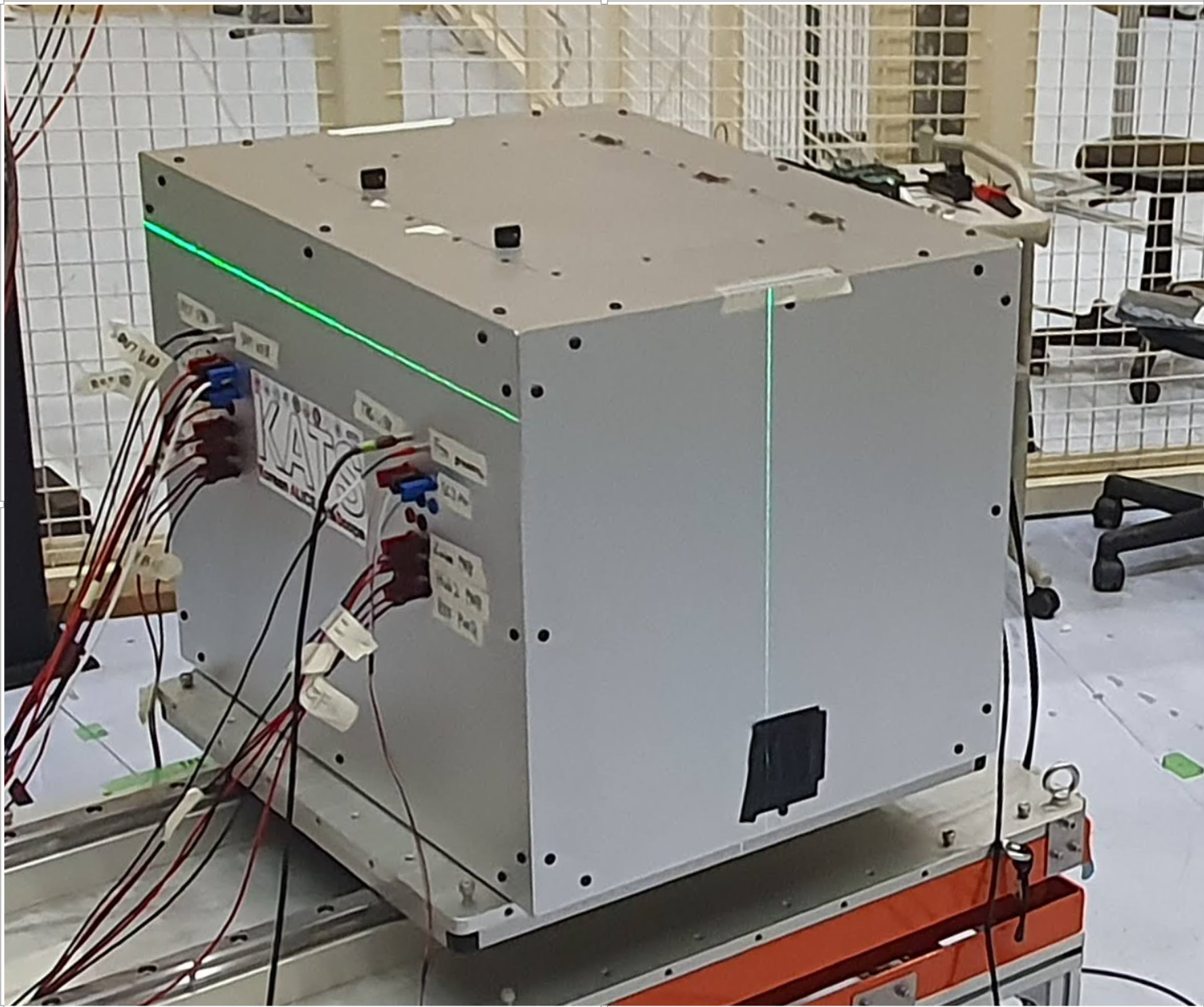}
    \caption{Schematic view of the telescope box layout (left) and photograph of the assembled telescope box (right).}
    \label{fig:telescopebox}
\end{figure} 

The telescope box, shown in Fig.~\ref{fig:telescopebox}, is a light-tight aluminum enclosure (approximately 674 $\times$ 524 $\times$ 504 mm$^{3}$, L $\times$ W $\times$ H) that houses all beam telescope components and shields the detector system from external light. Its modular structure, based on aluminum profiles and detachable rectangular plates, allows flexible reconfiguration to accommodate different DUT geometries and experimental layouts. The front and rear panels incorporate beam windows of 39 $\times$ 24 mm$^{2}$, providing unobstructed passage of the particle beam. One side panel is equipped with USB and BNC feed-through connectors for detector readout and trigger signal transmission, while the opposite panel hosts banana and LEMO feed-throughs for power distribution. Routing the readout and power cabling through opposite panels simplifies cable management and facilitates stable operation during beam tests.

An aluminum optical breadboard is mounted on the enclosure base, providing a stable mechanical reference for detector alignment. The breadboard features an array of 18 $\times$ 24 threaded holes on a 25 mm grid, enabling precise and reproducible mounting of the detector planes, trigger modules, and auxiliary components.

\section{Testbeam setup}
\subsection{Beam profile at KEK PF-AR}\label{Beam profile at KEK}

\begin{figure}[htb]
    \centering
    \includegraphics[width=0.54\textwidth]{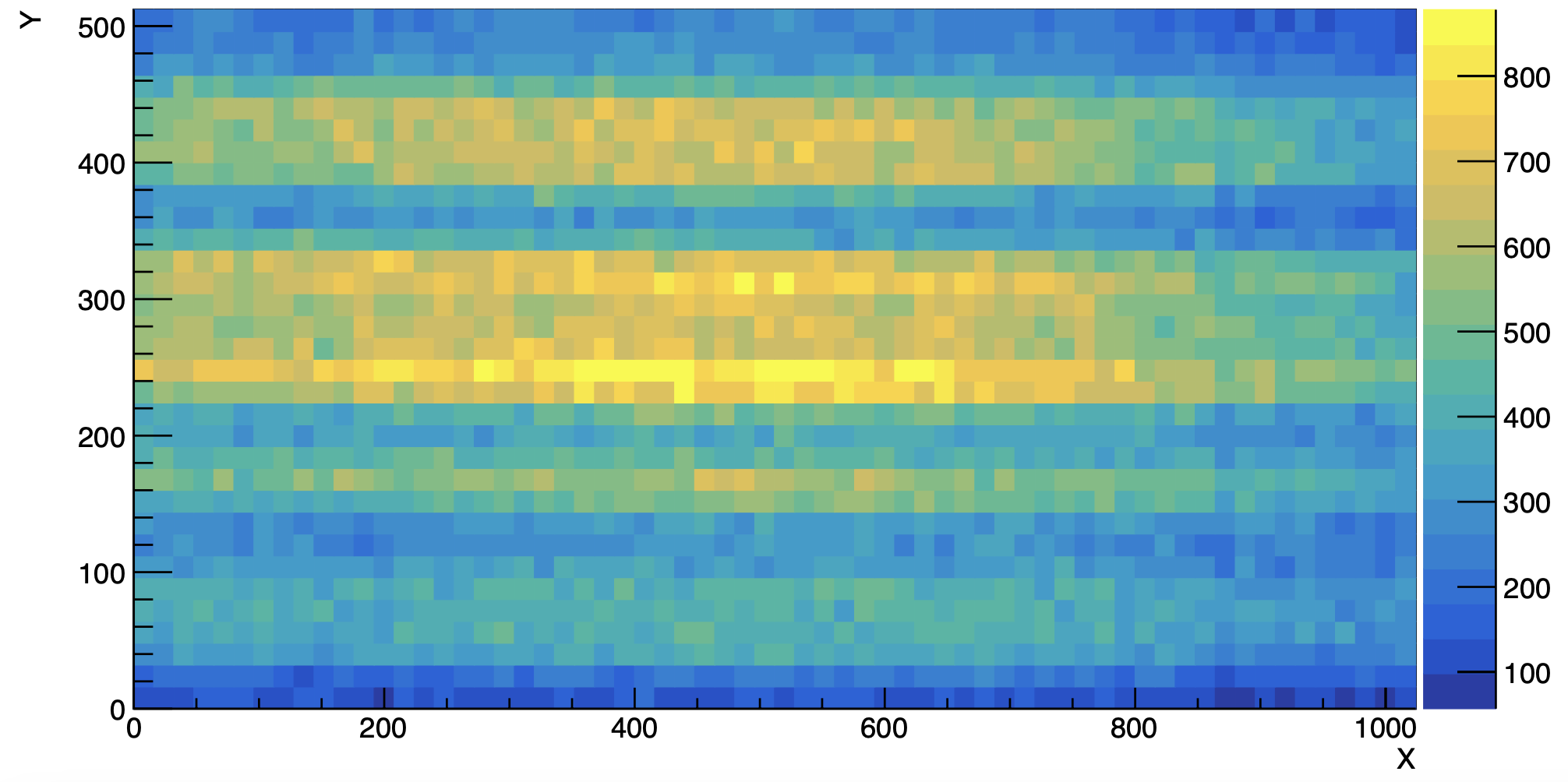}
    \includegraphics[width=0.40\textwidth]{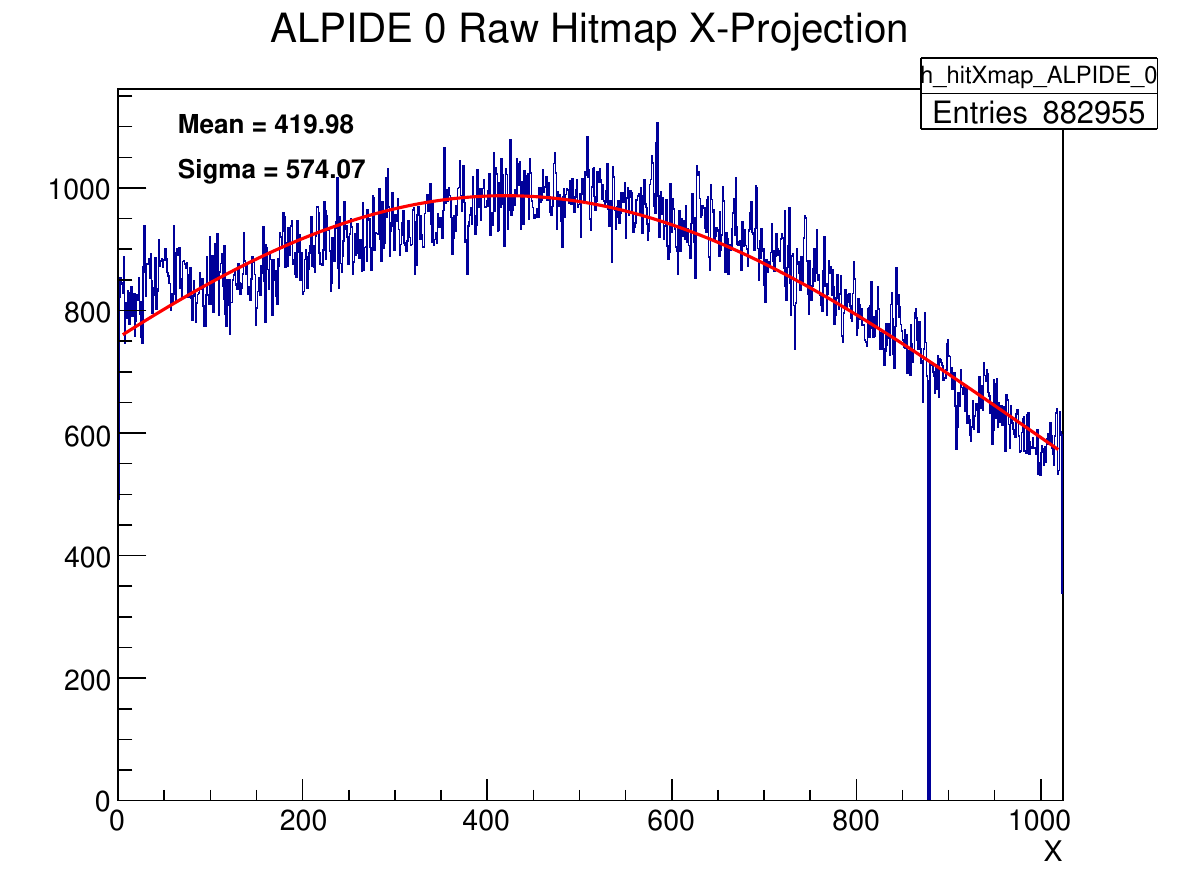}
    \caption{Rebinned hit map (left) and its projection onto the $x$ direction (right) for the ALPIDE 0 reference plane measured in the KEK PF-AR electron beam. The hit map was constructed by merging $8 \times 8$ original pixel bins along the $x$ and $y$ directions to better visualize the beam profile and spatial uniformity.}
    \label{fig:beaminfo}
\end{figure} 

The beam profile at the telescope entrance was characterized using the hit distribution recorded by the ALPIDE 0 reference plane, which is the first sensor encountered by the incoming beam. The hit map exhibits a characteristic horizontal striped pattern (left panel of Fig.~\ref{fig:beaminfo}). This pattern originates from local trigger inefficiencies in the scintillating-fiber modules: the 16 fibers of each module are arranged horizontally, and local efficiency is reduced both at the cladding between adjacent fibers and near the edges of the 4 $\times$ 4 fiber bundle, where the outer fibers overlap only partially with the 3 $\times$ 3 mm$^{2}$ active area of the MPPC.

We observed these trigger inefficiencies during initial commissioning with the AND coincidence trigger originally foreseen for the telescope. To minimize the impact of local inefficiencies in each module, we changed the trigger logic from AND to OR, so a valid trigger is issued when at least one of the two modules responds. This change was possible because the measured trigger rates in both modules were close to the beam particle rate, indicating that accidental trigger contamination was sufficiently low. To preserve the selection of particles traversing the full telescope, we placed both trigger modules downstream of all reference planes, ensuring that triggered tracks crossed every reference layer.

The striped inefficiency pattern extends along the horizontal fiber direction, so integrating over the vertical ($y$) axis averages out the stripe-to-stripe variation. The resulting $x$-projection of the hit map (right panel of Fig.~\ref{fig:beaminfo}) therefore provides an unbiased view of the beam distribution over the ALPIDE active area, and is well described by a Gaussian shape. This confirms that the beam uniformly covers the ALPIDE active area, consistent with the nominal PF-AR electron-beam spot size at the test-beam location of approximately 2 $\times$ 8 cm$^{2}$, sufficient to cover the entire 30 $\times$ 15 mm$^{2}$ ALPIDE active area.

\begin{figure}[htb]
    \centering
    \includegraphics[width=0.65\linewidth]{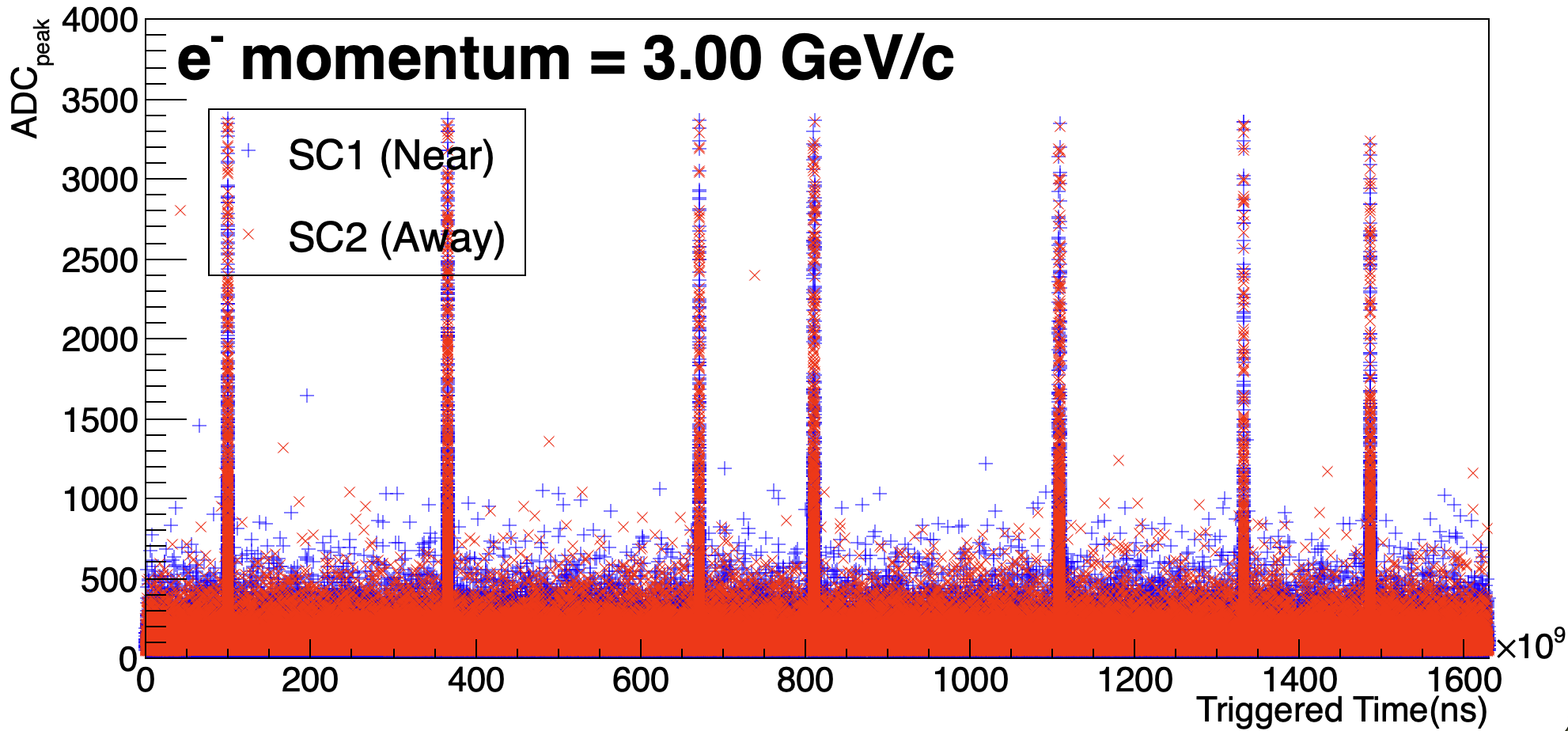}
    \includegraphics[width=0.65\linewidth]{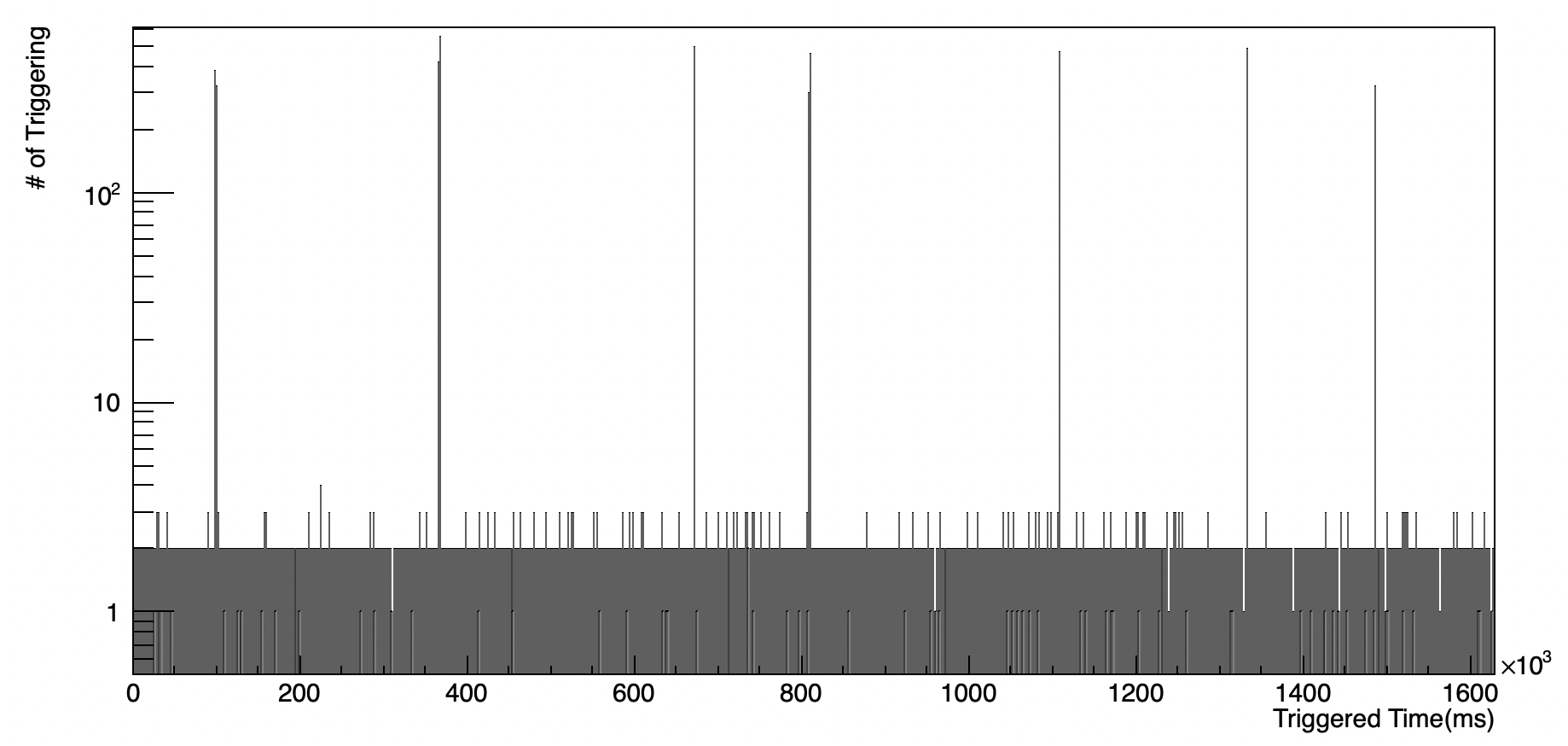}
    \caption{ Trigger time as a function of the pulse height (top), quantified by $\mathrm{ADC_{peak}}$, which is proportional to the energy deposited in the scintillating fiber. The peak ADC value is defined as the difference between the pulse's maximum ADC value and the pedestal, with a conversion factor of $1~\mathrm{ADC} = 0.6~\mathrm{mV}$. The trigger time is obtained by setting the first recorded event to $t = 0$ and converting the elapsed clock counts into physical time units. Distribution of the number of trigger events per 1~ms time interval for the same data sample (bottom). The vertical axis is logarithmic.}  
    \label{fig:timeinfo}
\end{figure}

The KEK PF-AR test beamline provided electron beams with tunable momenta between 0.5 and 5 GeV/$c$. We performed initial commissioning and alignment at 3 GeV/$c$, where beam intensity is highest and statistics are favorable. For the bent ALPIDE efficiency measurements presented in Section~\ref{Data analysis and results}, we used the highest available momentum, 5 GeV/$c$, to minimize scattering effects on track reconstruction.

The trigger and beam-structure analyses described below were performed with the 3 GeV/$c$ data. The MPPC analog signals were duplicated via a fan-in/fan-out module: one copy drove the hardware trigger logic of the telescope, while the other was sent to the FADC DAQ for waveform recording (right panel of Fig.~\ref{fig:trigger}). The FADC used an independent record trigger requiring both scintillator modules to simultaneously exceed a pulse-amplitude threshold of 50 ADC counts (1 ADC count = 0.6 mV), selecting clean beam-crossing events for offline analysis.

Figure~\ref{fig:timeinfo} shows the response of the trigger system during a representative data-taking period. The top panel displays the trigger time as a function of the pulse height ($\mathrm{ADC_{peak}}$), defined as the difference between the maximum ADC value of the pulse and the pedestal level; the trigger time is obtained by setting the first recorded event to $t = 0$ and converting the elapsed clock counts into physical time units. Signals from beam particles appear as a continuous baseline distributed around $\mathrm{ADC_{peak}}$ $\simeq$ 500 throughout the data-taking period. Superimposed on this baseline are pronounced spikes with saturated $\mathrm{ADC_{peak}}$ values, appearing at irregular time intervals of approximately 150–350 s. The bottom panel shows the same trend: the number of triggers per 1 ms interval on a logarithmic scale. During the spike periods, the instantaneous trigger rate exceeds the nominal operating rate by more than two orders of magnitude.

These features are attributed to the periodic beam-injection cycles of the KEK PF-AR, which maintain the circulating beam intensity. At PF-AR, the extracted electron beam is produced by gamma rays from the halo of the circulating electron beam interacting with a wire target, followed by electron-positron pair production in a copper converter~\cite{KEK_PFAR:2023smx}. During injection, the circulating beam is transiently perturbed, and the resulting changes in the beam properties alter the characteristics of the extracted particles reaching the test area, producing bursts of unusually energetic hits in the scintillator modules. These injection-related events are readily identified by their characteristic time structure and are excluded from the offline analysis by requiring single-track events.

\subsection{Data acquisition}

\begin{figure}[htb]
    \centering
    \includegraphics[width=0.9\linewidth]{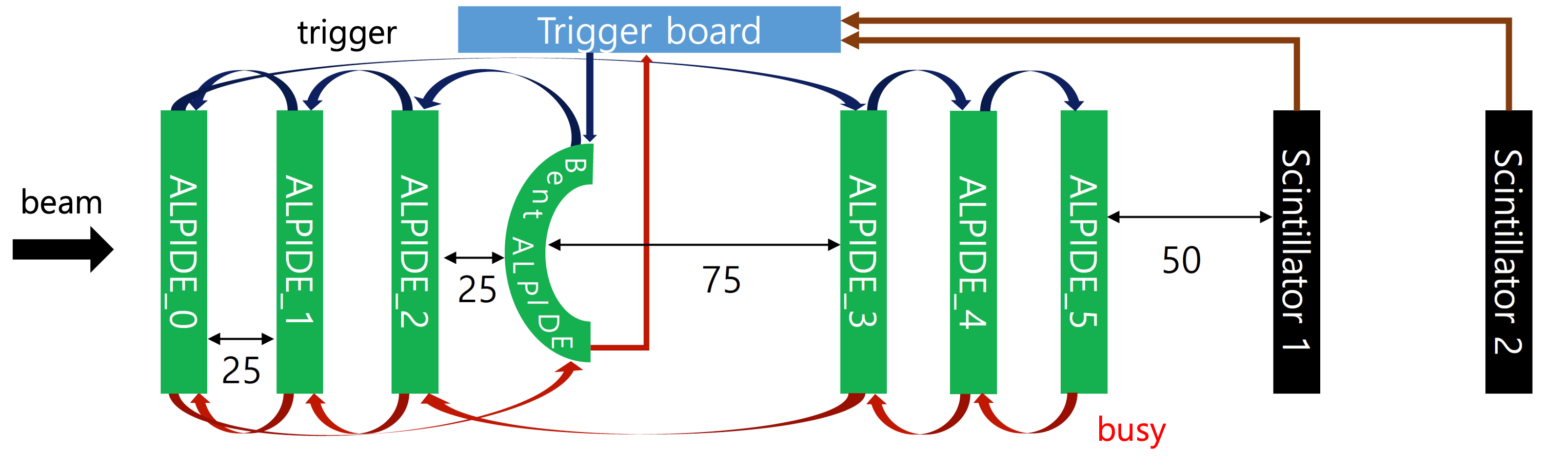}
    \caption{Schematic diagram of the DAQ chain of the Korea ALICE Telescope.}  
    \label{fig:daqchain}
\end{figure}

The DAQ chain of KATS is shown schematically in Fig.~\ref{fig:daqchain}. The scintillator trigger system (Section~\ref{subsec:trigger detector}) provides the primary trigger signal, which is distributed by a trigger board along the detector chain: the trigger is first received by the DUT and then sequentially propagates through ALPIDE 0 to ALPIDE 5, as indicated by the black arrows in the upper part of Fig.~\ref{fig:daqchain}. Busy signals, which indicate the readout status of each detector plane, propagate in the reverse order, so that the trigger board is aware of the readout status of every detector before issuing the next trigger. As long as any detector is busy, the system issues no new trigger, ensuring all planes remain synchronized during event building.

To avoid pile-up effects, the trigger logic enforces a minimum separation of 100 $\mu$s between consecutive triggers and a 50 $\mu$s busy-veto window after each trigger. The former prevents the same beam particle from generating multiple triggers, while the latter inhibits new triggers while a detector is still reading out the previous event. This protection is particularly relevant for the ALPIDE sensor, whose in-pixel amplifier pulse can extend over several tens of microseconds when operated at low thresholds~\cite{Kim:2016ktw}. The 100 $\mu$s separation corresponds to a sustained readout rate of up to 10 kHz, which is well above the measured beam particle flux of approximately 75 Hz/cm$^{2}$ (corresponding to about 340 Hz over the full ALPIDE active area) at the KEK PF-AR test beamline, so that pile-up effects were negligible throughout the beam-test campaign.

The DAQ software is based on the EUDAQ framework~\cite{Ahlburg:2019jyj, Liu:2019wim}, which provides synchronized readout of all detector planes, trigger information, and auxiliary data from the trigger board and environmental sensors (temperature and humidity) inside the telescope box. EUDAQ ensures coherent event building across the reference planes, DUT, and trigger system, enabling reliable offline reconstruction and performance analysis.

\section{Data analysis and results}\label{Data analysis and results}

\subsection{Data analysis flow}

\begin{figure}[htb]
    \centering
    \includegraphics[width=0.55\linewidth]{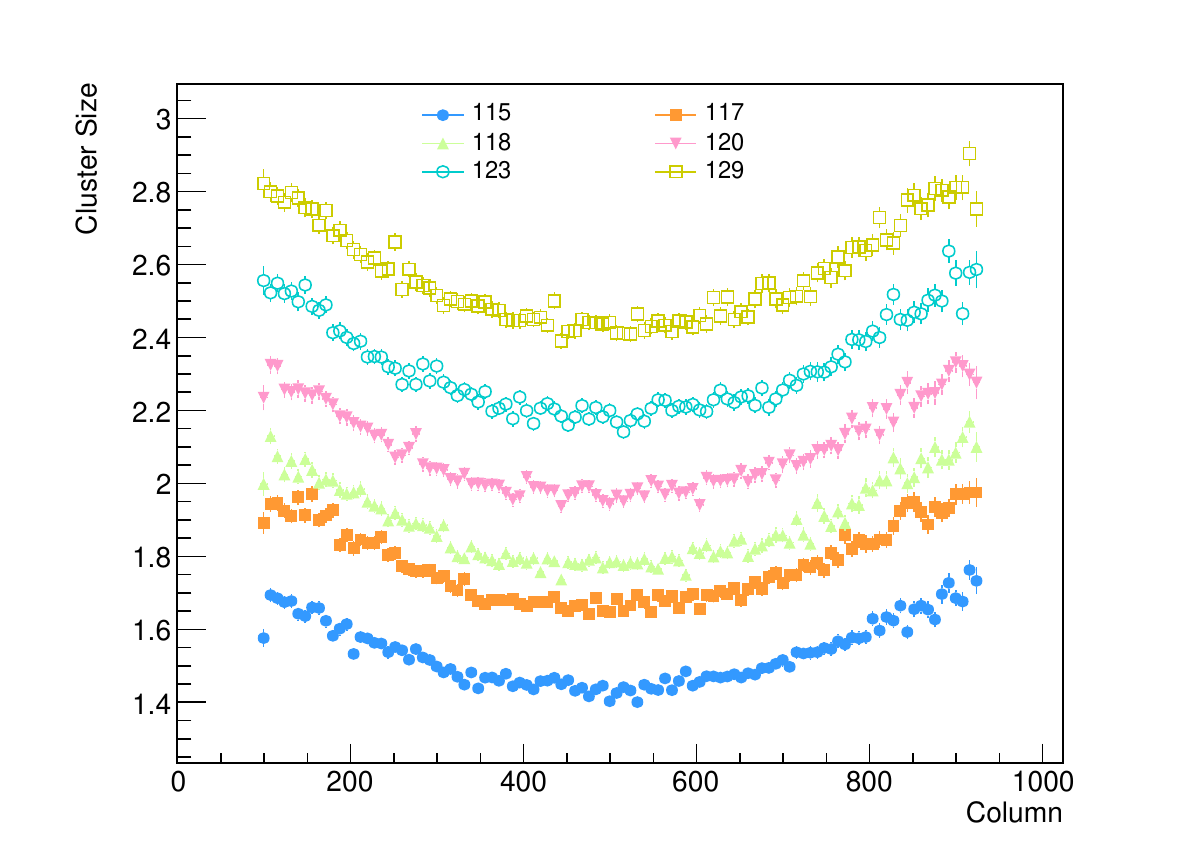}
    \caption{DUT cluster size as a function of column for different VCASN settings. Different marker colors represent different VCASN DAC values, corresponding to different discriminator threshold settings.}  
    \label{fig:trackAngleAndClusterSize}
\end{figure}

The data were processed using the Corryvreckan test-beam reconstruction framework~\cite{Dannheim:2020jlk}. We reconstructed particle tracks by fitting straight-line trajectories to clusters identified in the six reference planes and interpolating them to the $z$ position of the detector under test (DUT). Figure~\ref{fig:trackAngleAndClusterSize} shows the distribution of the reference-track angles. To ensure a clean, well-defined data sample, we applied strict event and track quality criteria. Only events containing exactly one reconstructed track were retained. In addition, tracks were required to satisfy a $\chi^{2}/\mathrm{ndf}<3$ condition and to have valid cluster hits on all reference planes.

\begin{figure}[htb]
    \centering
    \includegraphics[width=0.9\linewidth]{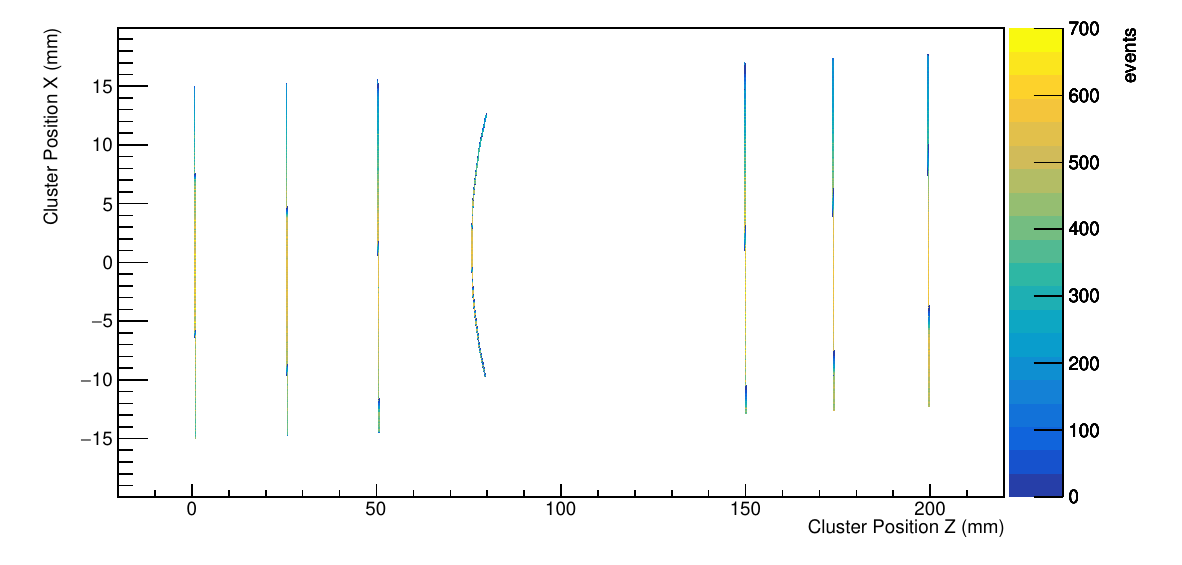}
    \caption{Distribution of global cluster positions in the $xz$ plane.}  
    \label{fig:clusterPositionsXZ}
\end{figure}

% Pre-pre-alignment
We performed detector alignment in several stages. First, we examined correlations between the cluster positions of the sensor planes. We observed a systematic shift in the mean cluster position in the $x$ direction that increased with layer ID. This trend is consistent with a global rotation of the telescope by approximately $0.7^{\circ}$ about the $y$ axis during the beam test. Accordingly, the $x$ and $z$ positions, as well as the orientation angles of the reference planes, were corrected in the geometry description.

%Pre-alignment, 
In the subsequent pre-alignment step, we corrected translational offsets in the $x$ and $y$ directions for all sensors, including the DUT, using the first ALPIDE layer as the reference plane. This was achieved by fitting the cluster-position correlations between the first layer and each of the remaining layers with Gaussian functions and applying the 
extracted mean shifts. For the DUT, the BentPixelDetector class~\cite{ALICE:2025_secondBent} determined global cluster positions from pixel coordinates while accounting for the bending radius and orientation. Figure~\ref{fig:clusterPositionsXZ} shows the reconstructed cluster positions in the $xz$ plane before the alignment procedure, illustrating the geometry of the bent DUT. The blue-colored regions indicate areas with relatively low hit occupancy. Furthermore, several complete pixel columns in the $y$ direction are inactive because of damaged sensor/readout channels, producing the vertical gaps observed in the hit map.

Figure~\ref{fig:trackAngleAndClusterSize} presents the DUT cluster size as a function of column for different VCASN settings. The different marker colors represent different VCASN values, with lower VCASN corresponding to higher discriminator thresholds. In the sensor center, the average cluster size varies from about 1.4 to 2.5 with threshold, showing the strong influence of the discriminator threshold on charge sharing and cluster formation. For all VCASN settings, the cluster size increases with distance from the detector center in the $x$ direction, reaching values approximately 20\% larger near the sensor edges. This behavior is attributed to the increasing local incidence angle across the curved sensor. For a bending radius of 18 mm, a position approximately 12 mm from the sensor center corresponds to a local incidence angle of about $40^\circ$, increasing the effective path length through the silicon by approximately 30\% relative to normal incidence. The observed increase in cluster size is therefore qualitatively consistent with the enhanced charge sharing associated with the longer path length through the sensor.

% Alignment of reference planes
Following pre-alignment, the reference planes were aligned using reconstructed tracks with hits on all six planes and a spatial matching window of $50~\mu\mathrm{m}$. 
We verified the alignment success by re-running tracking with the updated geometry. The resulting residual distributions exhibit widths of about $5~\mu\mathrm{m}$, consistent with the intrinsic spatial resolution of the ALPIDE sensor.

% DUT alignment and ROI
For DUT alignment, we kept the reference-plane geometry fixed and included the pre-aligned DUT in the tracking procedure. We applied an initial loose matching window of $300~\mu\mathrm{m}$ to associate DUT clusters. In addition, the outermost 100 pixels in the bending direction ($x$) were excluded from the analysis. Because the standard alignment procedure is less effective for curved sensors, we performed additional manual optimization steps. In particular, we iteratively varied the DUT rotation angles around the $x$ and $y$ axes and selected the optimal configuration based on the resulting residual distributions. After this procedure, residuals with a width of approximately $10~\mu\mathrm{m}$ and a mean below $1.5~\mu\mathrm{m}$ were obtained. 

\subsection{Detection efficiency}

\begin{figure}[htb]
    \centering
    \includegraphics[width=0.8\linewidth]{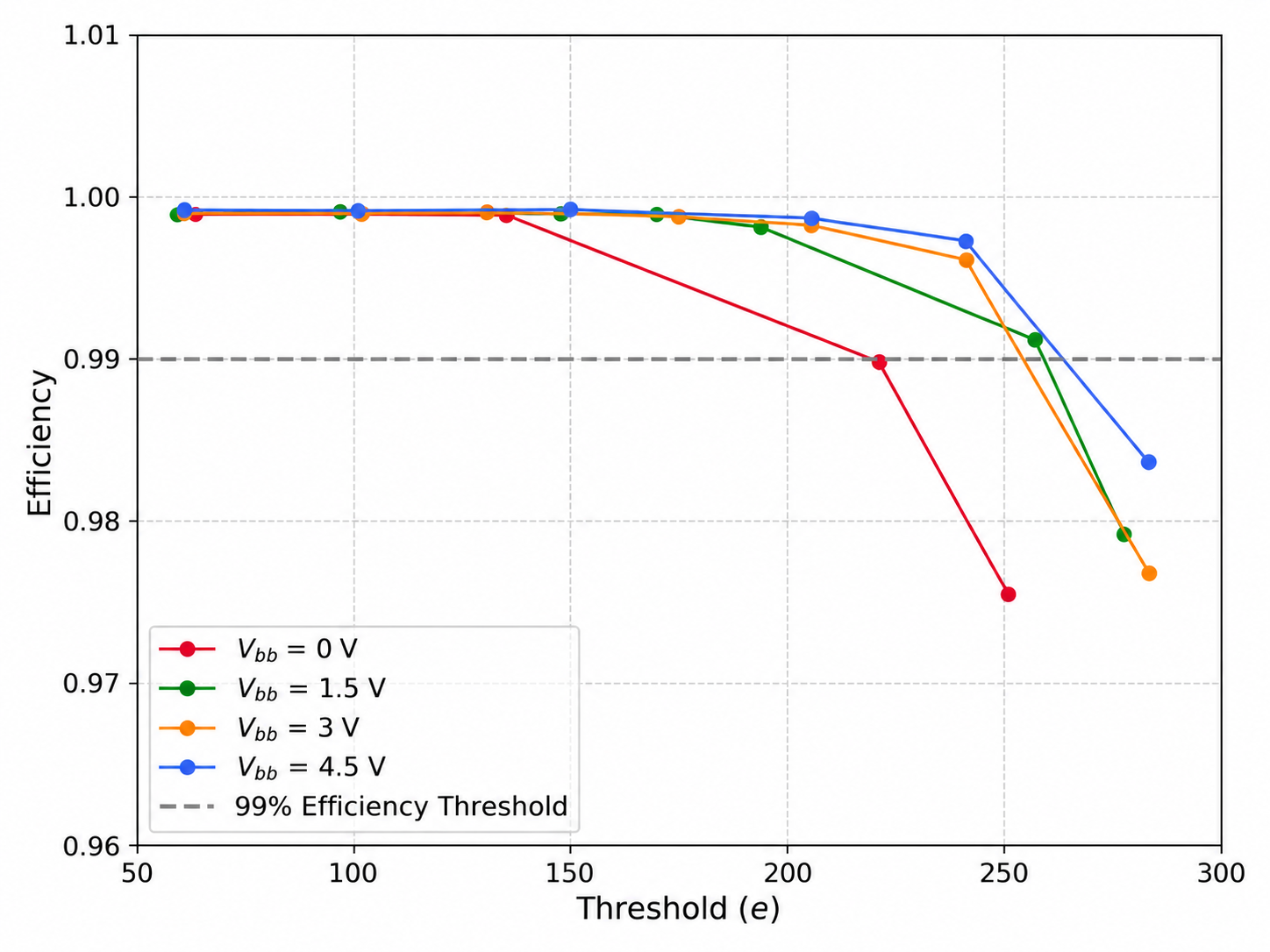}
    \caption{Detection efficiency of the DUT as a function of discriminator threshold for different reverse-bias voltage settings.}  
    \label{fig:effvsthreshold}
\end{figure}

Using the fully aligned geometry, DUT clusters located within $100~\mu\mathrm{m}$ of a reference track with $\chi^{2}/\mathrm{ndf}<3$ were associated with that track. The threshold values corresponding to each back-bias voltage setting, $V_{\mathrm{bb}}$, and VCASN configuration were determined in separate measurements. We then evaluated detection efficiency as a function of VCASN, i.e., the threshold. The results for all $V_{\mathrm{bb}}$ settings are shown in Fig.~\ref{fig:effvsthreshold}.

For $V_{\mathrm{bb}}=1.5$, 3.0, and 4.5~V, the efficiency exhibits a very similar dependence on threshold, remaining above 99\% up to a threshold of approximately 250~$e$. In contrast, without back-bias ($V_{\mathrm{bb}}=0$), the efficiency decreases more rapidly with increasing threshold, reaching 99\% at approximately 220~$e$. This behavior is consistent with results reported by the ALICE ITS3 collaboration~\cite{ALICE:2022_firstBent}.

\section{Summary}

In this work, we have developed and commissioned the Korea ALICE Telescope (KATS), a versatile silicon tracking telescope designed to support research and development of next-generation ultra-low-mass vertex detectors, with particular emphasis on technologies relevant to the ALICE ITS3 upgrade. The telescope integrates six ALPIDE Monolithic Active Pixel Sensors as high-precision reference planes, a bent ALPIDE sensor as the device under test, a scintillating-fiber-based trigger system, and a flexible, light-tight mechanical enclosure. The system operated successfully under high-energy electron-beam conditions at the KEK Photon Factory Advanced Ring (PF-AR).

The beam test demonstrated stable detector operation, reliable triggering, and robust data acquisition under realistic experimental conditions. Beam profile and timing-structure studies confirmed uniform illumination of the detector acceptance and clean event selection, despite injection-related rate bursts intrinsic to PF-AR operation. The EUDAQ-based data acquisition system enabled synchronized readout of all detector components, while the Corryvreckan framework provided a reliable reconstruction chain, including precise alignment of the telescope geometry. The achieved tracking performance, with residuals at a few micrometers for the reference planes, validates the overall detector performance and alignment procedure.

A key result of this study is the measurement of the bent ALPIDE sensor's detection efficiency. Within the statistical and systematic uncertainties, its performance is comparable to that of planar ALPIDE sensors operated under similar conditions. The efficiency remains above 99\% over a wide range of threshold and back-bias settings, demonstrating that bending to radii relevant for ITS3 does not degrade charge collection or hit-detection performance. These results agree well with previous ITS3 R\&D studies by the ALICE Collaboration, providing independent validation using a newly 
developed experimental platform.

In addition, the KATS enables precise characterization of beam properties and detector response, establishing a solid framework for future performance studies. Beyond its application at KEK PF-AR, the telescope's compact, modular design also makes it a promising tool for detector R\&D at domestic accelerator facilities such as RAON and KOMAC~\cite{Kwon2026KOMAC, Choi2025RAON, Park2026AcceleratorKorea}, where it could support beam characterization and silicon-sensor performance studies using available heavy-ion and proton beams.

Overall, the successful commissioning of the KATS and the demonstrated performance of the bent ALPIDE sensor confirm that the system is a reliable and powerful tool for advanced silicon detector R\&D. The KATS is well-suited for ongoing and future studies, including systematic investigations of spatial resolution, efficiency uniformity, alignment stability, and long-term operation of curved silicon sensors, thereby contributing to the development of next-generation tracking detectors for collider experiments.

\backmatter

%\bmhead{Supplementary information}

\bmhead{Acknowledgements}

We acknowledge support from the National Research Foundation of Korea (NRF) grant funded by the Korean government (MSIT) under Contract No. NRF-2008-00458 and Japan Society for the Promotion of Science (JSPS) KAKENHI Grant No. 24K00663. We thank the KEK PF-AR team for their valuable support during the beam test.

%\section*{Declarations}

%Some journals require declarations to be submitted in a standardized format. Please check the Instructions for Authors of the journal to which you are submitting to see if you need to complete this section. If yes, your manuscript must contain the following sections under the heading `Declarations':

%\begin{itemize}
%\item Conflict of interest/Competing interests (check journal-specific guidelines for which heading to use)
%\item Ethics approval and consent to participate
%\item Consent for publication
%\item Data availability 
%\end{itemize}


\bibliography{sn-bibliography}% common bib file

\end{document}